# Electroabsorption in MoS$_2$

D. Vella[1,2*], D. Ovchinnikov[3,4], N. Martino[5], V. Vega-Mayoral[1,2], D. Dumcenco[3,4], Y.-C. Kung[3,4], M.-R. Antognazza[6], A. Kis[3,4], G. Lanzani[6,7], D. Mihailovic[1,2,8,9], and C. Gadermaier[1,2*]

[1]Department of Complex Matter, Jozef Stefan Institute, Jamova 39, 1000 Ljubljana, Slovenia

[2]Jozef Stefan International Postgraduate School, Jamova 39, 1000 Ljubljana, Slovenia

[3]Electrical Engineering Institute, École Polytechnique Fédérale de Lausanne, CH-1015 Lausanne, Switzerland.

[4]Institute of Materials Science and Engineering, École Polytechnique Fédérale de Lausanne, CH-1015 Lausanne, Switzerland.

[5]Wellman Center for Photomedicine, Massachusetts General Hospital and Harvard Medical School, 65 Landsdowne St, Cambridge, MA 02139, USA

[6]Center for Nano Science and Technology, Italian Institute of Technology, Via Pascoli 70/3, 20133 Milano, Italy

[7]Department of Physics, Politecnico di Milano, P. Leonardo da Vinci 32, 20133 Milan, Italy

[8]Faculty of Mathematics and Physics, University of Ljubljana, Jadranska 19, 1000 Ljubljana, Slovenia

[9]Center of Excellence in Nanoscience and Nanotechnology, Jamova 39, 1000 Ljubljana, Slovenia

**To translate electrical into optical signals one uses the modulation of either the refractive index or the absorbance of a material by an electric field. Contemporary electroabsorption modulators (EAMs) employ the quantum confined Stark effect (QCSE)[1-3], the field-induced red-shift and broadening of the strong excitonic absorption resonances characteristic of low-dimensional semiconductor structures. Here we show an unprecedentedly strong transverse electroabsorption (EA) signal in a monolayer of the two-dimensional semiconductor $MoS_2$. The EA spectrum is dominated by an apparent linewidth broadening of around 15% at a modulated voltage of only $V_{pp}$ = 0.5 V. Contrary to the conventional QCSE, the signal increases linearly with the applied field strength and arises from a linear variation of the distance between the strongly overlapping exciton and trion resonances. The achievable modulation depths exceeding 0.1 dBnm$^{-1}$ bear the scope for extremely compact, ultrafast, energy-efficient EAMs for integrated photonics, including on-chip optical communication.**

Two-dimensional materials offer vast opportunities for photonics since they combine easy fabrication and mechanical flexibility with particularly strong light-matter interaction[4,5]. Semiconducting two-dimensional transition metal dichalcogenides[6,7] such as $MoS_2$ support tightly bound excitons[8] with particularly strong absorption cross sections, with a monolayer absorbing about 10% of the incident light (Figure 1a). This exceptional light-matter interaction is corroborated by cavity polaritons at room temperature[9], strong optical Stark effect[10], and ultrasensitive phototransistors[11].

Here we measure the transverse EA of monolayer $MoS_2$ by modulating the gate voltage in a transparent field effect transistor. The absorption spectrum of $MoS_2$ in Figure 1a shows the characteristic A and B exciton resonances[4-11]. The distance between the two main Raman peaks in Figure 1b, $\Delta\omega$ = 19.6 cm$^{-1}$ confirms the monolayer thickness[12]. The transistor is built in a bottom-gate configuration according to the structure in Ref. 13, using 30 nm of $HfO_2$ as gate dielectric and a sapphire slice as the transparent substrate (Fig 1c). We present data from two geometrically identical devices with different turn-on voltages (Figures 1d, 2b), which we ascribe to adsorption doping[14] since the measurements are carried out in air. We apply a modulated gate voltage (with peak-to-peak voltage $V_{pp}$) superimposed on a DC offset to the transistor and detect the change $\Delta T$ in the transmitted light intensity via a lock-in amplifier . Transmission is measured using a tuneable laser focussed on a <500 nm spot in a confocal microscope[15,16]. We normalise the modulated signal $\Delta T$ by the transmitted light intensity $T$, which directly yields the change $\Delta A$ in absorbance:

$$\Delta A = -\frac{\Delta T}{T} \tag{1}$$

The EA spectrum of monolayer $MoS_2$ for transverse field modulation consists of alternating positive and negative absorption changes (See Figures 2a-c), three peaks of increased absorption and two of decreased absorption. The two peaks of decreased absorption are stronger than the other three and peak at the same wavelengths as the A and B features in the absorption spectrum. For negative gate voltages (or $V_g$ < + 2V for device 1), where the transistor is in the *OFF* state, the shape and magnitude of the signal is independent of the DC component of the gate voltage. When the transistor is *ON* and charges are injected, the signal is weaker and slightly changes shape. The injected charges may partly screen the field and weaken the EA signal. Moreover, in this case the modulated field also modulates the charge concentration in the channel, which adds to $\Delta A$ (charge induced absorption)[17,18].

The most common features of EA are changes in peak positions (red- or blue-shifts), peak width (usually broadening), and in some cases also changes in the area under the absorbance curve in a certain spectral window (transfer of oscillator strength between different electronic transitions)[1,2]. In the $\Delta A$ spectra these signal components appear with distinctive shapes related to the respective absorption peaks: the same as the absorption peaks' for transfer of oscillator strength, first derivative for shifts, and second derivative shape for changes in width. We fitted two Gaussian peaks and a generic exponential function to account for the background from various scattering mechanisms to the absorption spectrum in Fig 1a and used the derivatives of these Gaussians independently to fit the EA spectra, as shown in Figures 2d,e. We find broadening (about 15% for the A peak of device 2) as the main effect, accompanied by a slight red shift. The joint action of broadening and red shift slightly slants the two negative peaks towards higher energies compared to the absorption resonances and diminishes the middle peak due to the overlap of positive and negative signal components.

For different modulation voltages $V_{pp}$, the signal retains its shape and increases linearly in magnitude (Figures 2c,f). This means that in particular the broadening, as the dominant signal component, increases linearly with the modulated field, a behaviour that has not been reported in other materials so far. On the other hand, shifts with linear field dependence are common and indicate built-in fields or permanent dipole moments[19]. In engineered two-dimensional semiconductor structures (quantum wells or multiple quantum wells), the QCSE yields a red shift proportional to the square of the applied electric field[1,2], accompanied by a broadening of the excitonic resonances. The broadening is ascribed to an increased homogeneous linewidth due to field-induced exciton dissociation (charge transfer into the "walls" of the quantum well), and has a highly non-linear field dependence. Hence our results disagree with the conventional QCSE.

To explain the linear field dependence of the broadening, we take a closer look at the linewidth of the absorption resonances. Since $MoS_2$ from any synthesis method is doped[13,20], the absorption peak A consists of at least two overlapping electronic resonances, $A^0$ from the neutral ground state to a neutral exciton state, and $A^-$ from a charged ground state to a charged exciton state (trion)[17]. Similarly, the broader B peak should also be composed of two peaks $B^0$ and $B^-$. Although B trions have not been studied in detail yet, they play a role in the transient absorption spectra observed in femtosecond pump-probe experiments[21]. Since positive and negative trions appear at almost the same spectral position[22], the same considerations can also be made for *p* doping. The distance between $A^0$ and $A^-$ changes almost linearly with $V_g$ over a wide range of gate voltages comprising both the *ON* and *OFF* states[18, 22]. This shift of both features has been tentatively ascribed to Stark effect[22] or to a renormalization of both the band gap and the exciton (or trion) binding energy[18]. The latter explanation is applicable only for the *ON* state of the transistor, where the gate voltage actually modulates the charge density. Here, in the *OFF* state the charge density in the $MoS_2$ channel stays constant; hence we continue to discuss our results in terms of Stark effect.

By adding the field-induced absorbance change $\Delta A$ to the absorbance *A* without field in Figure 2g (to reduce noise, we used the fit for *A* rather than the original data) we obtain the shape of the absorbance spectrum in the presence of an electric field corresponding to a $V_g = V_{pp} = 0.5$ V. Clearly, the A peak is lowered and broadened and its composition by two peaks is now more clearly visible. A similar but smaller effect is also seen for peak B, which should also be composed of two resonances $B^0$ and $B^-$. However, peak B is too wide to yield accurate parameters for two component peaks; hence we do not explore its composition further. In Figs 2h and 2i we show fits of *A* and $A+\Delta A$ using two Gaussians for peak A, with height and width unchanged between both cases, only different positions. This fit embodies a model where the $A^0$ and $A^-$ resonances maintain their width and oscillator strength, but change their positions as a consequence of the applied electric field. A gate voltage of $V_g = 0.5$ V increases the distance between the $A^0$ and $A^-$ peaks from 36 to 48 meV, resulting in both a broadening and lowering of the A peak as a whole. Additionally, the peak shifts slightly to the red since the red shift of the $A^-$ peak is stronger (-9 meV) than the blue shift of the $A^0$ peak (+3 meV). While the composite A peak with 36 meV between the $A^0$ and $A^-$ peaks (Figure 2i) still looks very similar to a Gaussian, at 48 meV it does not (Figure 2h), which explains why the fit with first and second derivatives of a Gaussian deviates from the real data (Figure 2e).

The gate voltage $V_g$ causes a transverse electric field in the MoS$_2$ monolayer

$$F = \frac{V_{gs}}{d_M + \frac{\varepsilon_M}{\varepsilon_H} d_H} \qquad (2)$$

With the indices $M$ and $H$ indicating the thickness $d$ and dielectric constant $\varepsilon$ of MoS$_2$ ($\varepsilon_M$ = 7) and HfO$_2$, ($\varepsilon_H$ = 19) respectively. A gate voltage $V_g$ = 0.5 V yields a field $F$ = 400 kVcm$^{-1}$. Our field modulation corresponds to a voltage difference of 20 V in the device of Ref. 18, giving a comparable shift of several meV for both the A$^0$ and the A$^-$ peak. A Stark shift has been recently reported for the low-temperature photoluminescence of a MoS$_2$ monolayer sandwiched between two dielectrics[23]. The A peak position follows a hysteretic curve, which the authors ascribe to a joint effect of Stark effect and charge transfer with the top dielectric (Al$_2$O$_3$). Along most of their curve, our field modulation of 400 kVcm$^{-1}$ would give a shift of a few meV in their device, consistent with our results. Please note that also in their data the linewidth changes, although they discuss only the peak shift.

A linear shift in the applied field $F$ points towards a permanent dipole moment or an induced dipole from a built-in electric field:

$$\Delta E = \overrightarrow{\Delta p} \cdot \vec{F} \qquad (3)$$

With $\Delta p$ being the difference in the dipole moment of the exciton (trion) and the neutral (charged) ground state. The shifts of a few meV correspond to differences in dipole moments between excited and ground states of a few Debye. Note that this is only a rough estimate obtained from the fits in Figs 2h,i.

The MoS$_2$ monolayer itself should have no out-of-plane dipole moment for symmetry reasons, but the asymmetry of the device (HfO$_2$ on one side of the monolayer, metal electrodes and air on the other) can induce such dipoles, which are specific to the device structure and its environment, and not an intrinsic property of MoS$_2$.

The dipole moment induced by the device structure can be manipulated by adsorbed molecules, since our experiments have been carried out in air. Adsorbed molecules can also dope the channel[14] and thus change the ratio between the A$^0$ and A$^-$ peaks. Device 1 (Fig 1d) is *p* doped, as confirmed by the higher turn-on voltage. This lowers the A$^-$ peak or even replaces it by an A$^+$ peak. A lower A$^-$ peak means a lower EA signal. An A$^+$ peak has never been reported in MoS$_2$, however, in WSe$_2$, compared to A$^-$ it shows a smaller shift with field[24], in the opposite direction, which again lowers the EA. In Figure 3 we show transmission and EA maps of both devices at the strongest EA peak (1.9 eV). Although we average over a spot size of <500 nm, we find that the signal between the two electrodes (a lower signal is also seen below the electrodes, probably due to multiple reflections in the substrate) varies between 0.15 % and 0.22 % for Device 1 and 0.40 % and 0.52 % for Device 2.

Two key figures for an electromodulating device are the insertion loss *IL* (the transmission in the *ON* state of the EAM, not to be confused with the *ON* state of the transistor) and the modulation depth *MD* (the ratio between the transmission in the ON and OFF states). Assuming we directly use the present device as an EAM in transmission geometry, the two figures simply add up logarithmically to the optical density *OD* of the monolayer:

$$IL + MD = OD \qquad (4)$$

We have shown switching from approximately 89% transmission to 89.5% transmission, corresponding to an insertion loss of 0.48 dB and a modulation depth of 0.02 dB. Note that graphene, the strongest electromodulator to date[25], with the change in absorption caused by changes in the occupation of electronic levels rather than Stark effect, has an intrinsic limit of 0.10 dB for the modulation depth (2.3% transmission modulation[26]).

In conclusion, we have shown the electroabsorption effect in monolayer $MoS_2$ and ascribed it to linear field-induced shifts of the exciton and trion resonance peaks in opposite directions. The strength of the effect suggests that 2d semiconductors are excellent materials for EAMs. . Compared to graphene, we envisage a possibly even shorter device with much lower power consumption due to the low off current[13]. Additionally, since the working principle is not based on changes in the population of levels, there is no excited state lifetime limit to the response time[25]. Monolayer $MoS_2$ shows its strongest EA signal at 1.9 eV (650 nm). However, there is a vast choice of 2D materials[27,28] with bandgaps covering the whole range from 0 to beyond 5 eV, including the main telecom wavelengths (850-1550 nm). The achievable modulation depth exceeding 0.1 dBnm$^{-1}$ makes this concept also ideally suited for use directly in transmission geometry, which would require an optical path through the active material of only 10-50 nm. . Such a device could be extremely compact and enable on-chip optical communication.

Acknowledgements: We are grateful for the inspiring discussions with S. Bertolazzi, M. Caironi, V.V. Kabanov, D. Neumaier, V. Perebeinos, and J. Strle. Device fabrication was carried out in the EPFL Center for Micro/Nanotechnology (CMI). We thank Z. Benes (CMI) for technical support with e-beam lithography. The research leading to these results has received funding from the Marie-Curie ITN "MoWSeS" (grant no. 317451),the European Union's Seventh Framework Programme FP7/2007-2013 under Grant Agreement No. 318804 (SNM) and the Swiss SNF Sinergia Grant no. 147607. We acknowledge funding by the EC under the Graphene Flagship (grant agreement no. 604391).

Figure captions

**Figure 1 Optical and electrical characterisation of MoS$_2$ monolayer transistors. a,** extinction spectrum of the MoS$_2$ monolayer. **b**, Raman spectrum of the MoS$_2$ monolayer. **c**, schematic of the transistor geometry. **d**, dependence of the current ($I_{ds}$) of the gate voltage ($V_g$) for Device 1.

**Figure 2 Electroabsorption spectra of MoS$_2$ monolayer transistors. a,b,** electroabsorption spectra of devices 1 and 2 for different gate voltage DC offsets. Inset in b shows the gate voltage dependence of the current in device 2. **c**, electroabsorption spectra of device 1 for different peak-to-peak voltage modulation amplitude $V_{pp}$. **d,e** normalised electroabsorption spectrum (dotted) of device 1 and first and second derivatives of the absorption peaks used to fit the spectra (individual contributions in d, cumulative fit in e). **f,** A and B peak amplitudes from **c** as a function of $V_{pp}$. **g**, fit of the extinction spectrum (blue), electroabsorption spectrum (red) and sum of the two (magenta) give the extinction spectrum at a gate voltage $V_g = V_{pp}$. **h,i**, extinction spectra with and without field fitted to three Gaussian curves.

**Figure 3 electroabsorption and transmission maps of MoS$_2$ monolayer transistors. a,b,** electroabsorption and transmission maps of device 1. **c,d,** electroabsorption and transmission maps of device 2. All measurements were performed at 650 nm, electroabsorption with a DC offset of -1V and a modulated voltage $V_{pp}$ = 0.5 V.

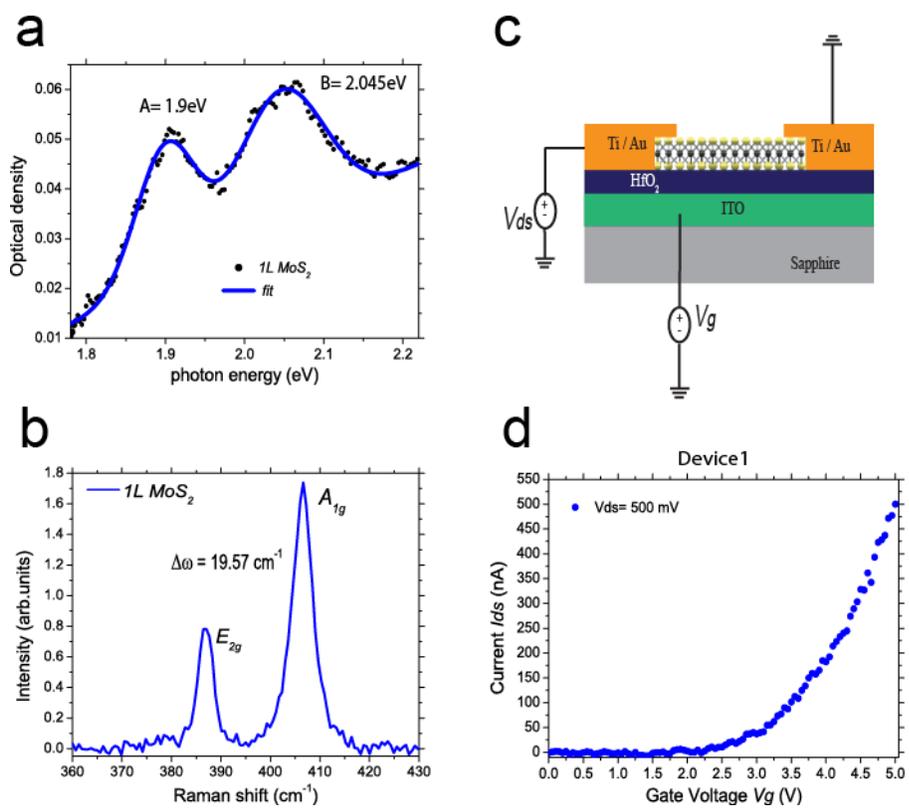

Fig. 1

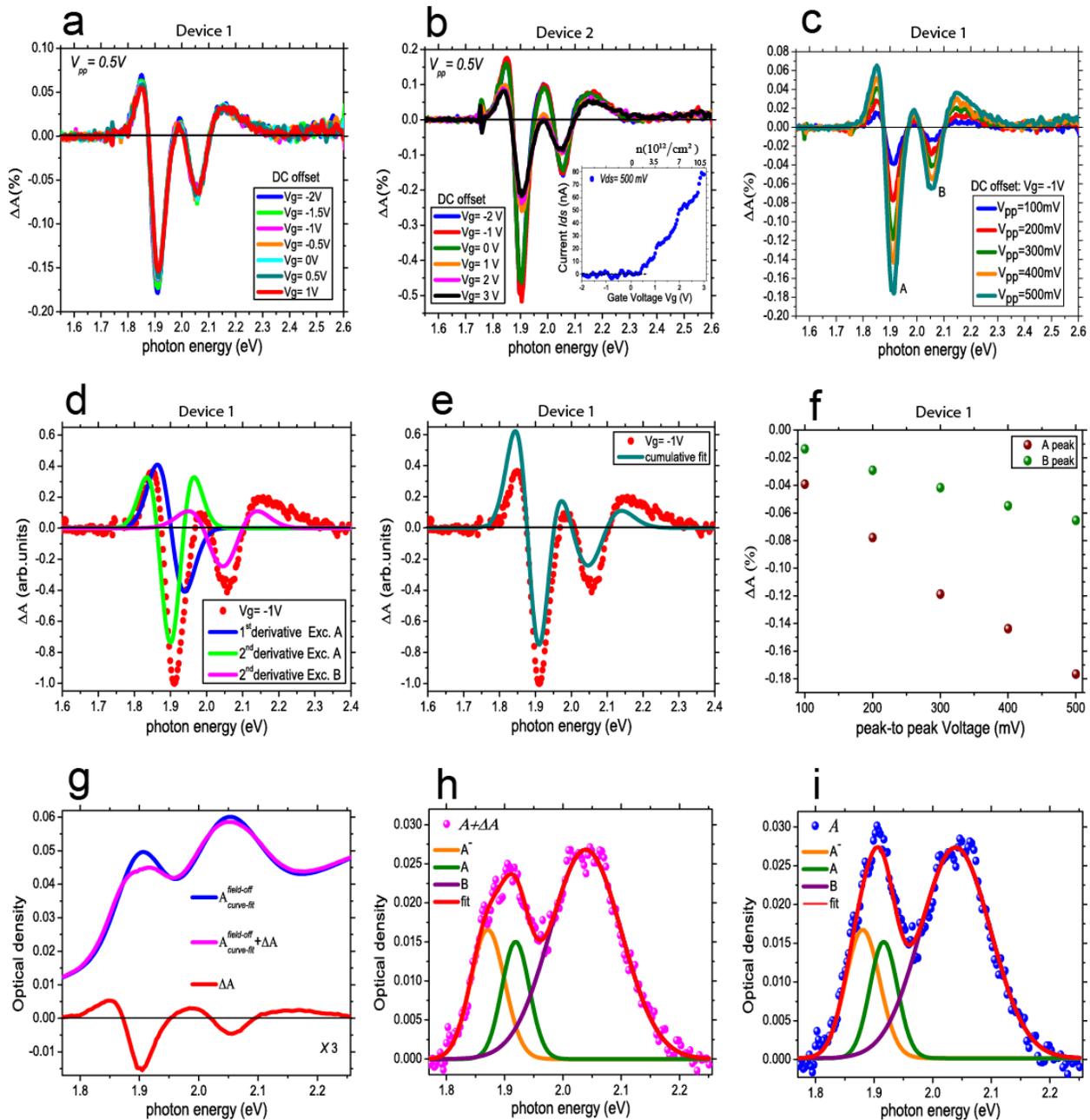

Fig. 2

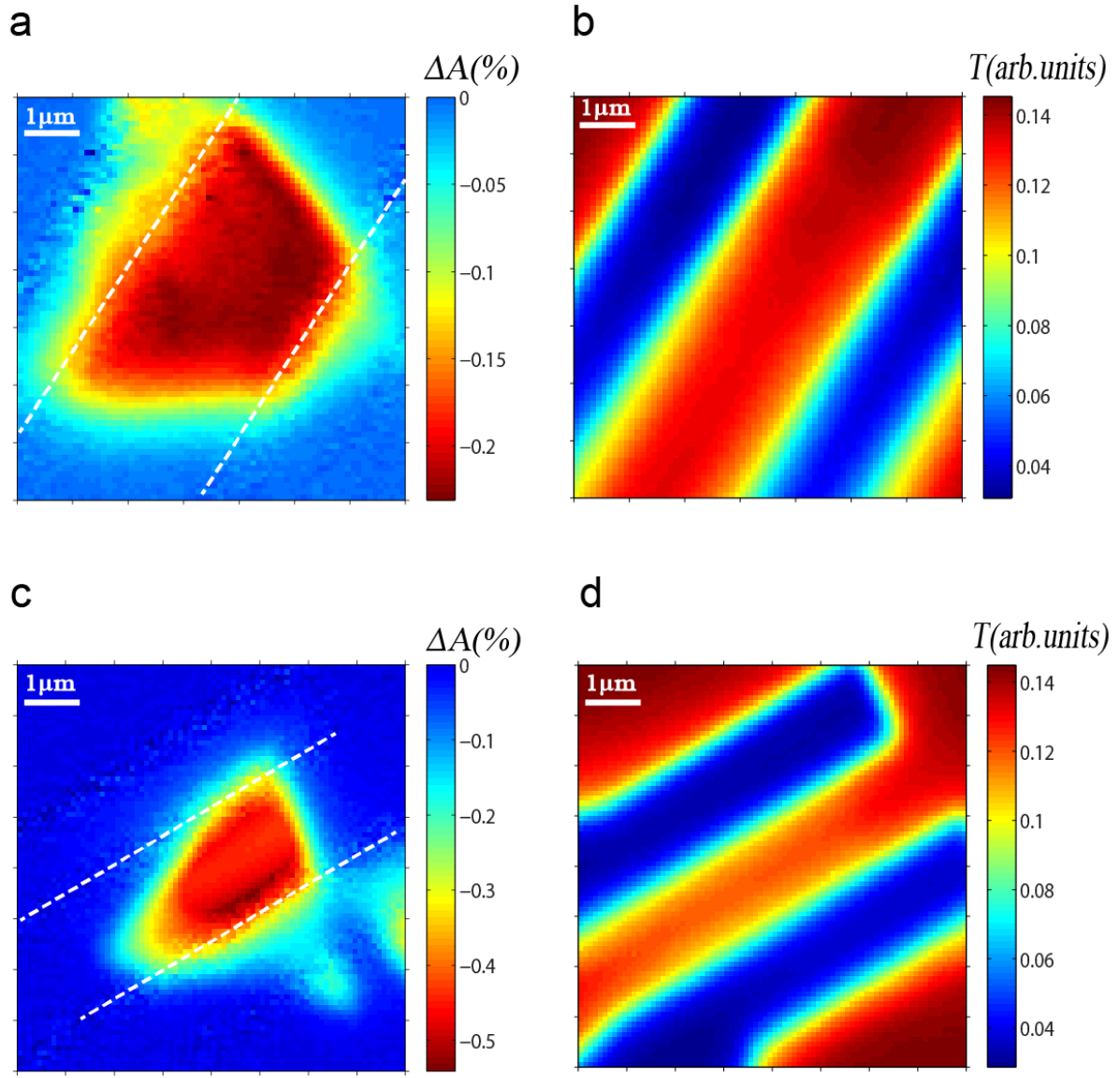

Fig. 3